\documentclass[journal=jacsat,manuscript=article]{achemso}
\setkeys{acs}{articletitle = true}

\usepackage[version=3]{mhchem} 

\usepackage{textcomp}
\usepackage{subcaption}
\usepackage[locale=DE]{siunitx}

\author{Alexandra Göbel}
\affiliation{Max Planck Institute for the Structure and Dynamics of Matter, Center for Free Electron Laser Science, Hamburg, Germany.}
\author{Angel Rubio}
\affiliation{Max Planck Institute for the Structure and Dynamics of Matter, Center for Free Electron Laser Science, Hamburg, Germany.}
\alsoaffiliation{Nano-Bio Spectroscopy Group and European Spectroscopy Facility (ETSF), Universidad del País Vasco CFM CSIC-UPV/EHU-MPC \& DIPC, 20018 Donostia-San Sebastián, Spain.}
\alsoaffiliation{Center for COmputational Quantum Physics, Simons Foundation Flatiron Institute, New York, NY, USA.}
\author{Johannes Lischner}
\affiliation{Department of Materials, Imperial College London, London SW7 2AZ, UK. The Thomas Young Centre for Theory and Simulation of Materials, London SW7 2AZ, UK.}
\email{alexandra.goebel@mpsd.mpg.de,   angel.rubio@mpsd.mpg.de, j.lischner@imperial.ac.uk}

\title{Light Induced Charge Transfer from Transition-metal Doped Aluminium Clusters to Carbon Dioxide}

\abbreviations{}
\begin{document}

\begin{abstract}
Charge transfer between molecules and catalysts plays a critical role in determining the efficiency and yield of photo-chemical  catalytic processes. In this paper, we study light-induced electron transfer between transition metal doped aluminium clusters and \ce{CO2} molecules using first-principles time-dependent density-functional theory. Specifically, we carry out calculations for a range of dopants (Zr, Mn, Fe, Ru, Co, Ni and Cu) and find that the resulting systems fall into two categories: Cu- and Fe-doped clusters exhibit no ground state charge transfer, weak \ce{CO2} adsorption and light-induced electron transfer into the \ce{CO2}. In all other systems, we observe ground state electron transfer into the \ce{CO2} resulting in strong adsorption and predominantly light-induced electron back-transfer from the \ce{CO2} into the cluster. These findings pave the way towards a rational design of atomically precise aluminium photo-catalysts.

\end{abstract}

\section{Introduction}

Transformation of \ce{CO2} into useful chemicals is currently one of the most important environmental and societal global challenges and there is significant interest in discovering novel catalysts which facilitate such reactions in an environmentally friendly setup.\cite{chang2016co, remiro2019photoassisted, rodemerck2013catalyst, kumar2012photochemical} A key intermediate step in this process is the transfer of an electron into the lowest unoccupied molecular orbital (LUMO) of the \ce{CO2} which results in chemical activation.\cite{alvarez2017co2} This can be achieved either thermally\cite{kho2017review}, for example when the molecule is adsorbed at an electron-donating metal site, electro-chemically\cite{hussin2019recent}, via plasma\cite{puliyalil2018review} or photo-chemically\cite{chang2016co}, i.e. through illumination by light.

Homogeneous catalysts, such as \ce{[Rh(cod)Cl2]-Ph2P(CH2)4PPh2}\cite{LeitnerRhCO2formic_acid} or  ``FCAT" \cite{costentin2014ultraefficient_homogenous_cat} have been used to transform \ce{CO2} into useful products, such as formic acid or carbon monoxide. Despite their impressive turn-over frequencies (TOFs) \cite{filonenko2014mechanism}, homogeneous catalysts are usually expensive to synthesize\cite{wang2011_hydrogenation_co2} and separation of the catalyst from the reaction products can be difficult. In addition, homogeneous catalysts can suffer from degradation which can lead to very short life-times.\cite{costentin2014ultraefficient_homogenous_cat} In contrast, heterogeneous catalysts are typically cheaper and easier to fabricate, but exhibit lower yields and are less selective.\cite{wang2011_hydrogenation_co2, prasad2008fischer} Common heterogeneous catalytic systems used in \ce{CO2} conversion are Fe based catalysts like Fe-Cu-Al-K(2) which exhibits \ce{CO2} conversion efficiencies of up to 30~\% \cite{prasad2008fischer}, Cu/ZnO catalysts\cite{nakamura2003issue} with 15\% CO yield and supported Ni catalysts with \ce{CO2} conversion yields of almost 50~\%\cite{chang1998hydrogenation}.

Atomically defined catalysts have the potential to overcome the challenges described above. Such catalysts consist of well-defined nanostructures composed of a specific number of atoms. They are highly controllable and allow for a detailed understanding of the relationship between structure and activity.\cite{li2013atomically} For example, \ce{[Cu32H20(S2P(OiPr)2)12]} has been shown to reduce \ce{CO2} to \ce{CO}.\cite{tang2017lattice} A lot of research has been focused on atomically defined gold structures\cite{jin2020toward, li2013atomically, zhu2011catalysis}. Despite the need for further stabilization, e.g. through ligands or immobilization onto surfaces, the atomically defined \ce{Au_n(SR)_m} systems have been shown to be robust towards thermal degradation at temperatures up to 200~\textdegree C.\cite{li2013atomically} For example, \ce{[Au25(SR)18]-} was shown to be catalytically active for the electroreduction of \ce{CO2}\cite{kauffman2014probing}. In addition, the structural effect of doping of this cluster was studied by Fei et. al. in atomically defined \ce{Au24M(SR)18} (M = Pt, Pd, Hg, and Cd) in which they could show that Hg and Cd are always located in an icosahedron shell position and never at the cluster core.\cite{fei2019metal}

Aluminium nanostructures have received significant attention in recent years for applications in photocatalysis because of their attractive and highly tunable optical properties.\cite{knight2014aluminum, gerard2014aluminium, martin2014fabrication} In addition, aluminium is relatively cheap, has one of the highest natural abundances and is easy to process\cite{moscatelli2012plasmonics, clavero2014plasmon}. Due to aluminium's non-precious character it is very prone to oxidization.\cite{knight2014aluminum} The formed oxide layer, though, has been shown to be porous and allows for diffusion of small molecules.\cite{gonzalez1999characterization} For example, Halas and coworkers performed reverse-water-gas-shift reactions on Al@CuO nanocrystals \cite{robatjazi2017plasmon}. Halas and coworkers also showed that aluminium nano-crystals can be decorated with different transition-metals (Fe, Co, Ni, Ru, Rh, Pd, Ir and Pt) to increase light absorption and generate hot-spots, making these systems highly interesting for various applications like plasmonic photo-catalysis or surface enhanced spectroscopies.\cite{Halas2017transition_metal_Al}

To gain a mechanistic understanding of catalytic processes and the effect of dopants, the theoretical modelling of the atomic and electronic structure of (photo-)catalysts is useful. For the case of Al nanostructures, Zhao et al. used first-principles density-functional theory (DFT) to understand the effect of introducing dopant atoms and additional charge on the adsorption of \ce{CO2}.\cite{zhao2013adsorption} They focused on small icosahedral clusters containing 12 Al atoms and one dopant (located at the center of the icosahedron) and found that these systems can activate \ce{CO2}.\cite{zhao2014theoretical} However, they only studied ground-state properties of these systems and did not investigate electronic excitations of relevance to photocatalysis.

In this paper, we study charge transfer between transition metal doped aluminium clusters and a \ce{CO2} molecule in both the ground state and the excited state. In particular, we carry out first-principles DFT calculations to determine the relaxed atomic geometry of \ce{[Al12M]-CO2} systems (with M denoting Zr, Mn, Fe, Ru, Co, Ni or Cu) and analyze the ground state charge transfer between cluster and \ce{CO2}. Except for the Fe and Cu doped systems, we observe significant electron transfer into the \ce{CO2} which results in chemical activation. Next, we carry out time-dependent density-functional theory (TDDFT) calculations to study the light-induced charge transfer in these systems. For the Fe and Cu doped cluster, we observe light-induced electron transfer into the \ce{CO2}. In contrast, systems with significant ground state charge transfer predominantly exhibit a back-transfer of electrons to the cluster in the excited states. Interestingly, the Zr and Co doped clusters have light-induced charge transfer in both directions.

\section {Methods}

We study the atomic and electronic properties of small transition-metal atom doped aluminium clusters with adsorbed \ce{CO2} molecules. In particular, we start from icosahedral \ce{[Al13]} and replace one of the outer Al atoms by one of the following atoms: Fe, Mn, Co, Cu, Ni, Ru or Zr. The resulting system is denoted by \ce{\textbf{[}Al12\textit{M}\textbf{]}} with \textit{M} being the transition metal. 

Next, we place a \ce{CO2} molecule in the vicinity of the transition metal atom and relax the atomic positions using density-functional theory (DFT). We have found that the relaxed structures depend sensitively on the details of the computational approach, including choice of basis, exchange-correlation functional and spin-polarization. To identify low-energy configurations, we first create a pool of candidate structures in the following way: we start by placing a linear \ce{CO2} molecule at a distance of \SI{2.7}{\angstrom} from the dopant atom with its axis perpendicular to the direction from the dopant to the cluster center. For this starting configuration, we first carry out a relaxation with spin-unpolarized DFT using the PBE exchange-correlation functional\cite{perdew1996generalized} with a Grimme D2 correction\cite{grimme2010consistent} to capture van-der-Waals interactions. The resulting structures were then used as starting points for a relaxation with spin-polarized DFT. These calculations were performed using the Quantum Espresso software package\cite{giannozzi2009quantum}, ultra-soft GBRV pseudopotentials\cite{garrity2014pseudopotentials}, a wave function cutoff of 40~Ry and a charge density cutoff of 200 Ry. The cluster was placed in a cubic box with linear dimensions of \SI{45}{\angstrom} and a Gaussian smearing of 0.001 Ry was employed. The relaxations were stopped when the forces were converged to within $10^{-3}$~eV/\si{\angstrom}. To test the accuracy of these pseudopotential calculations, we also carried out all-electron relaxations using the FHIaims code\cite{blum2009ab, havu2009efficient}. In these calculations, we also used the PBE exchange-correlation functional and second tier basis functions of numerical localized orbitals. Finally, for each cluster, the lowest-energy atomic structure was determined by comparing the total energies of all relaxed candidate structures in all possible spin states. For this, we used the local density approximation (LDA) and norm-conserving Hartwigsen- Goedecker-Hutter pseudopotentials\cite{hartwigsen1998relativistic} (including semicore states for the transition metal atoms) as implemented in the Octopus code\cite{castro2006octopus, andrade2015real, tancogne2020octopus, tancogne2017self} (which was also used for the time-dependent DFT calculations of excited-state properties). In the Octopus code, the Kohn-Sham wavefunctions are calculated on a real-space grid. The grid spacing and box size were converged for each system, see Appendix.

For the lowest-energy relaxed structures, excited-state properties were obtained using time-dependent density-functional theory (TDDFT). Specifically, absorption spectra were calculated using a real-time propagation of the time-dependent Kohn-Sham orbitals following a delta-function perturbation. In these calculations, the time step is chosen sufficiently small to ensure a high degree of total energy conservation over the total propagation time of 13.164~fs. All convergence parameters used are listed in the Appendix.

To analyze light-induced charge transfer between cluster and \ce{CO2} molecule, we also carry out linear-response TDDFT calculations. For this, we solve the Casida equation\cite{casidaReview} which has the form of an eigenvalue problem. The eigenvalues $\omega_I$ have the interpretation of excited state energies. Casida proposed that the eigenvectors $F_{I,cv}$ can be used to construct an approximate wavefunction of the excited state.\cite{casidaReview} Specifically, $C_{I,cv}=\sqrt{(\epsilon_c-\epsilon_v)/\omega_I} F_{I,cv}$ denotes the coefficient of the singly excited Slater determinant $|\Psi_{cv} \rangle=a^\dagger_c a_v |\Psi_0 \rangle$ in the excited state $I$, with $\epsilon_c$ and $\epsilon_v$ being the Kohn-Sham energies of unoccupied and occupied states, respectively. These quantities can be used to calculate the light-induced charge transfer ({LICT}) in the excited state $I$ as
\newline 
\begin{equation}
    \textit{LICT}_I = f_I (Q_0 - Q_I) = f_I \sum_{cv} (q_{c} - q_{v}) |C_{I,cv}|^2,
\end{equation}
where $Q_I$ and $Q_0$ denote the amount of charge localized on the \ce{CO2} molecule in the excited state $I$ and the ground state, respectively, and $f_I$ is the oscillator strength.  Also, $q_c$ and $q_v$ are the amount of charge localized on the molecule in the unoccupied conduction $c$ and occupied valence $v$ Kohn-Sham orbitals obtained by projecting the Kohn-Sham wavefunctions onto the atomic orbitals of the \ce{CO2} molecule. Note that a positive value of the \textit{LICT} indicates that electrons are transferred from the cluster onto the \ce{CO2} molecule. We note that LICT is often analyzed using the transition densities of the excited state.\cite{zong2018photoinduced, plasser2012analysis, sun2008visualizations} We have compared the LICT from our approach with the one obtained from transition densities for several excited states and found good qualitative agreement between the two approaches. To converge the optical spectra from the linear-response approach, 100 unoccupied states in each spin channel were included.

\section{Results}

Table~\ref{gsdata geometry} summarizes the results of the geometry relaxations of the transition-metal doped Al clusters with \ce{CO2} molecules. In all clusters except \ce{[Al12Zr]} the relaxed structures are distorted icosahedra. The distortion is caused by the transition-metal dopant (and not by the presence of the \ce{CO2} molecule). The almost perfect icosahedral geometry of \ce{[Al12Zr]} can be explained by its valence electron number which is 40 (not counting the 8 semi-core electrons). This is equal to the number of electrons in \ce{[Al13]-} which is also referred to as a "magic" cluster due to its closed shell, perfect icosahedral geometry and comparably high stability.\cite{smith2011electron}

Adsorption distances, defined as the smallest distance between a cluster atom and an atom of the \ce{CO2}, range from \SI{1.8}{\angstrom} to \SI{2.4}{\angstrom} for the doped clusters. In comparison, the adsorption distance between \ce{CO2} and the un-doped neutral \ce{[Al13]} cluster is significantly larger with \SI{3.6}{\angstrom}. The smallest adsorption distances are found for \ce{[Al12Mn]-CO2}
, \ce{[Al12Co]-CO2} and \ce{[Al12Zr]-CO2}, while the biggest ones occur for \ce{[Al12Fe]-CO2} and \ce{[Al12Cu]-CO2}. Except in \ce{[Al12Zr]-CO2} (where the \ce{CO2} is closest to one of the Al atoms neighboring the Zr), the most favorable adsorption site is the transition metal dopant.

For all clusters except \ce{[Al12Fe]} and \ce{[Al12Cu]}, the bond angle of the adsorbed \ce{CO2} differs significantly from the value of the isolated molecule which is 180$^\circ$. The deviation from linearity increases as the adsorption distance is reduced. For example, \ce{[Al12Co]-CO2} with an adsorption distance of \SI{1.8}{\angstrom} has a bond-angle of  135.5$^\circ$, while \ce{[Al12Ru]-CO2} with an adsorption distance of \SI{2.1}{\angstrom} has a bond-angle of 142.9$^\circ$. The bending of \ce{CO2} reduces the energy of the lowest unoccupied energy levels and therefore facilitates transfer of electrons onto the \ce{CO2}.\cite{alvarez2017co2} This is often referred to as catalytic activation of the \ce{CO2} molecule.

\begin{table}
\begin{tabular}{ p{2.5cm} p{2.1cm} p{2.1cm} | p{2.5cm} p{2.1cm} p{2.1cm}}
 \hline
 System & Adsorption distance [\si{\angstrom}] & O-C-O \mbox{angle [{\textdegree}]} & System & Adsorption distance [\si{\angstrom}] & O-C-O \mbox{angle [{\textdegree}]}\\
 \hline
 &&&&&\\
 & \includegraphics[width=19mm,scale=0.08]{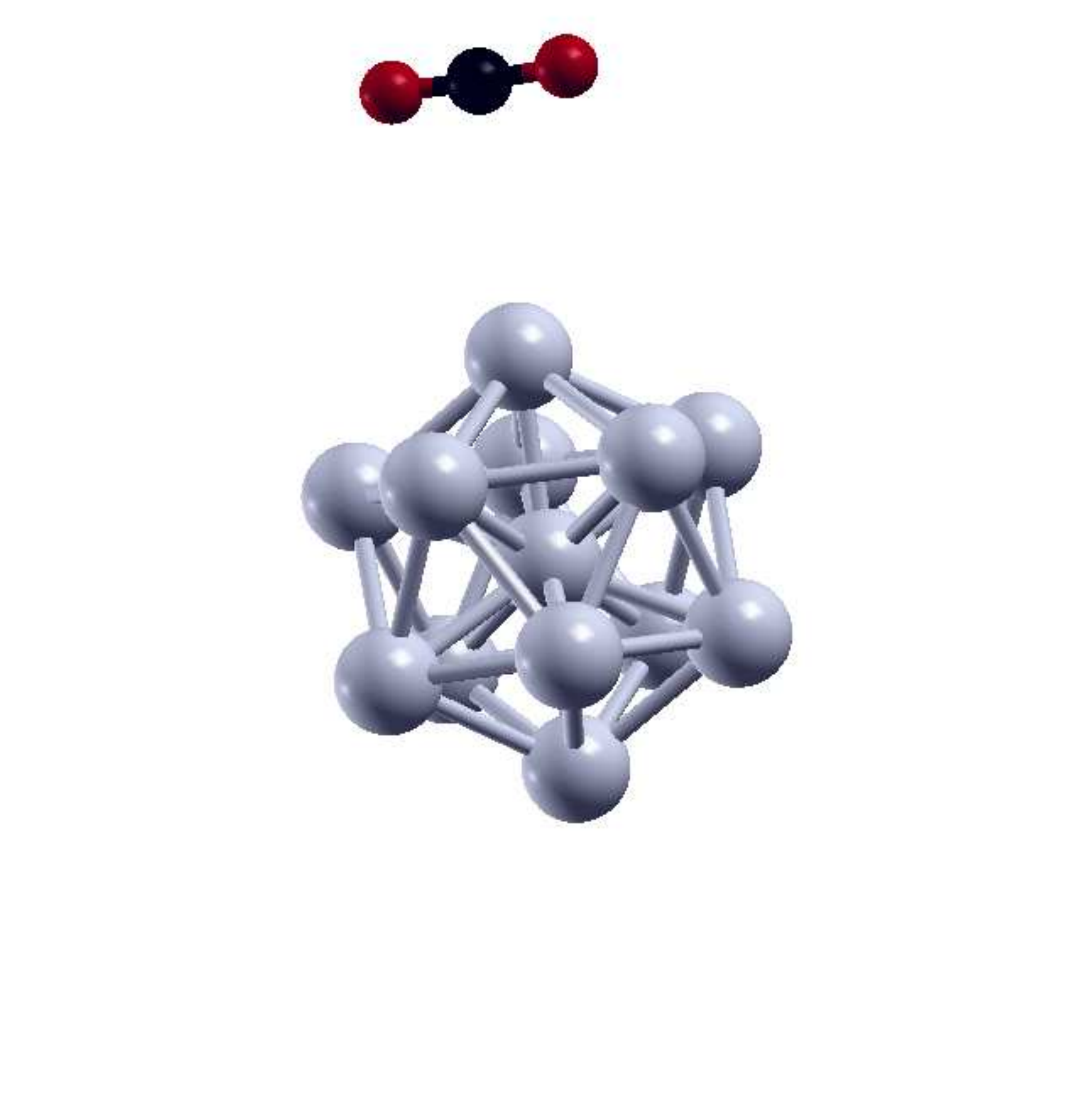} & & & \includegraphics[width=19mm,scale=0.08]{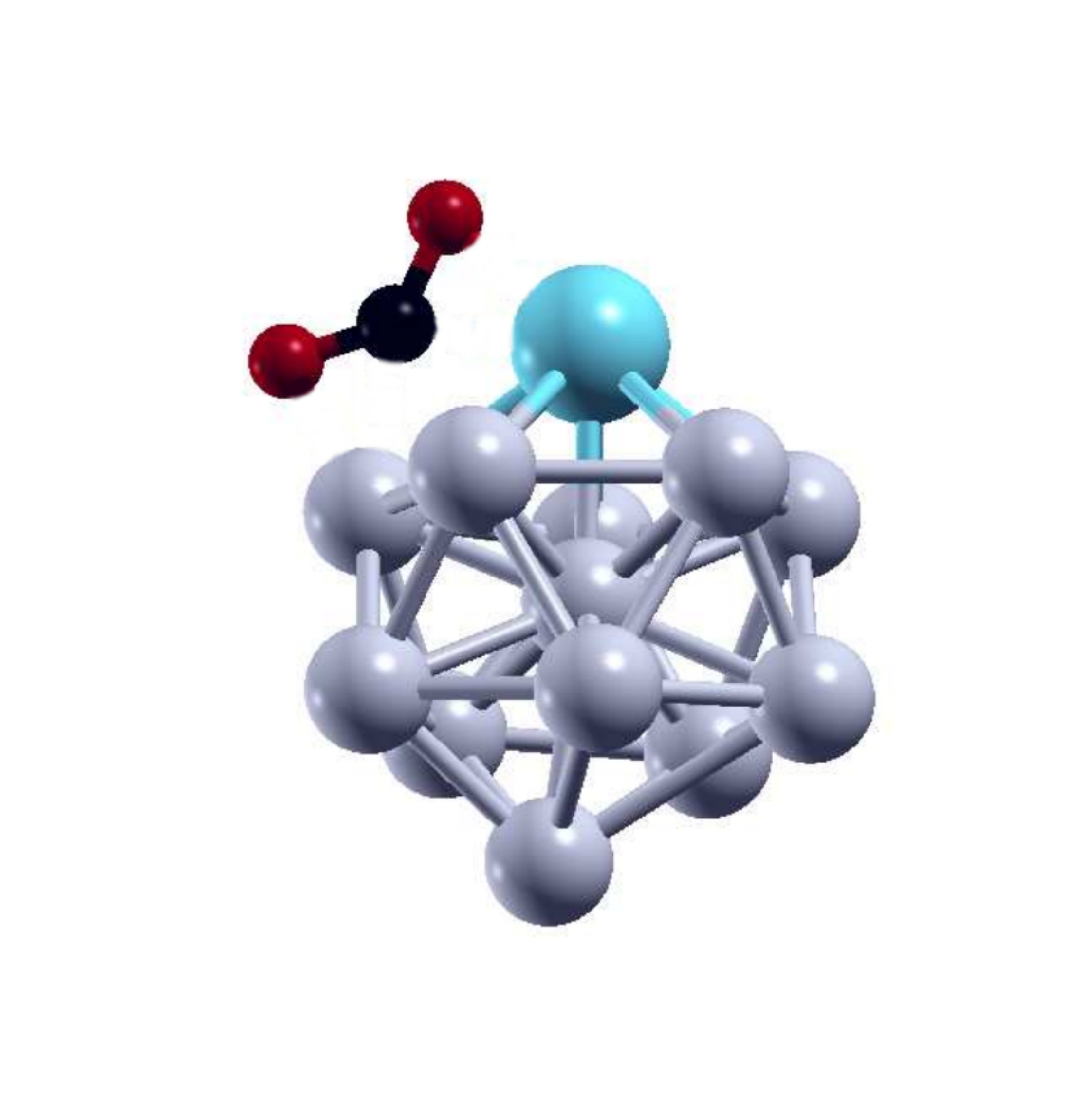} &\\\relax
 &&&&&\\
 \ce{[Al13]-CO2} & 3.6 (C-Al) & 177.7 & \ce{[Al12Zr]-CO2} & 1.9 (O-Al) & 131.4 \\
 \hline
 &&&&&\\
 & \includegraphics[width=19mm,scale=0.08]{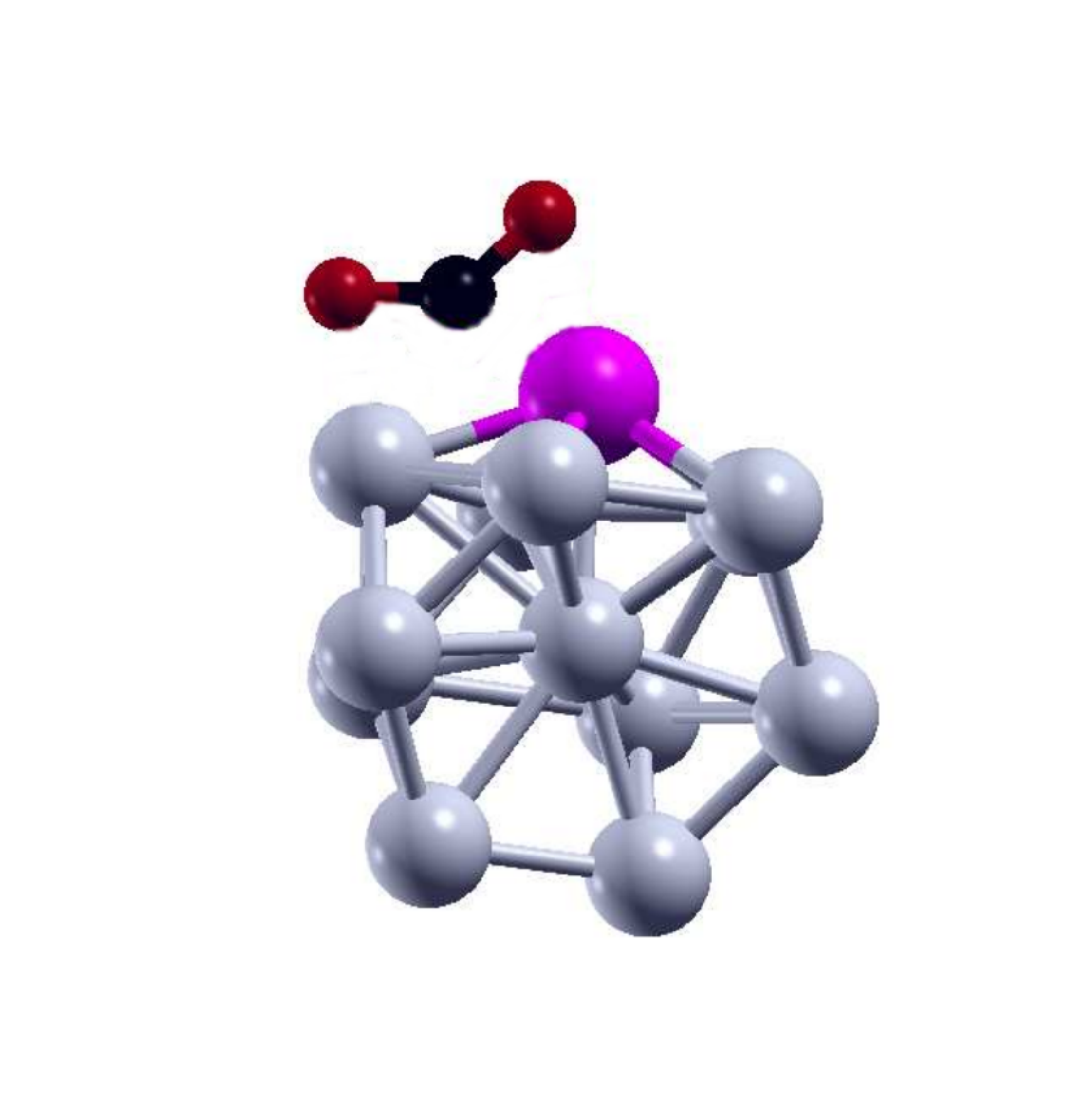} &&& \includegraphics[width=19mm,scale=0.08]{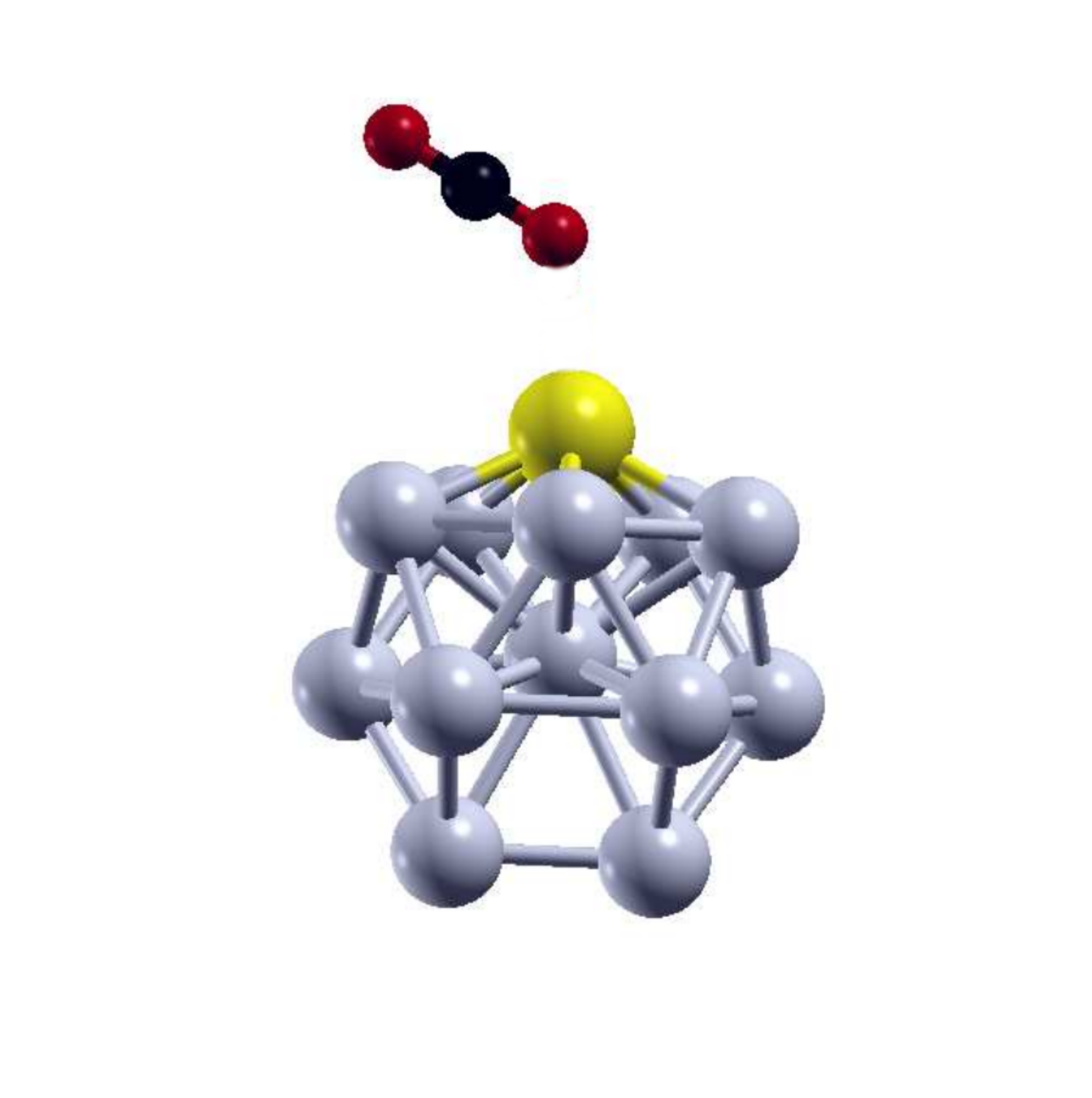} & \\\relax
 &&&&&\\
 \ce{[Al12Mn]-CO2} & 1.8 (C-Mn) & 137.1 & \ce{[Al12Fe]-CO2} & 2.4 (O-Fe) & 179.7 \\
 \hline
 &&&&&\\
 & \includegraphics[width=19mm,scale=0.08]{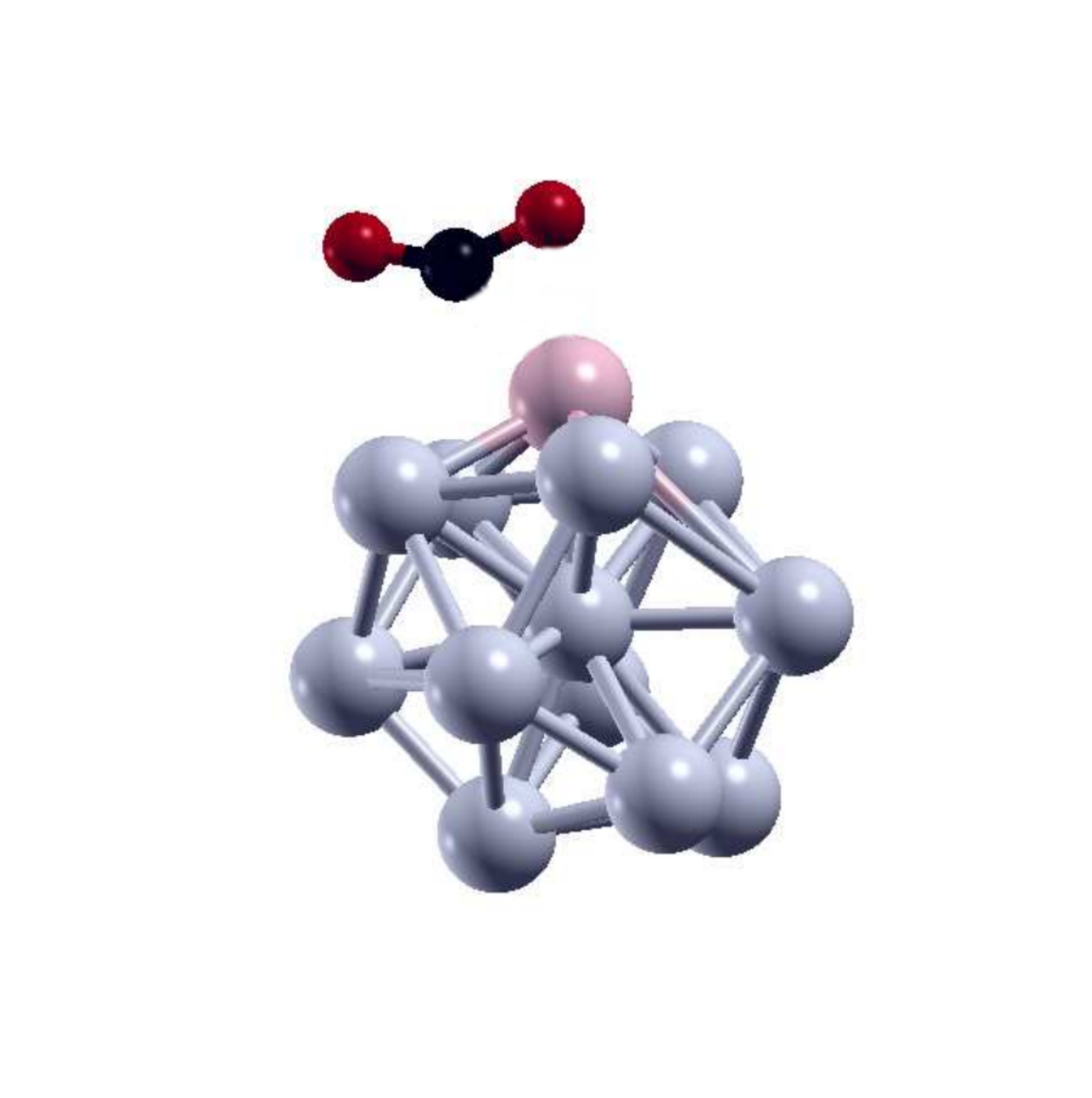} & && \includegraphics[width=19mm,scale=0.08]{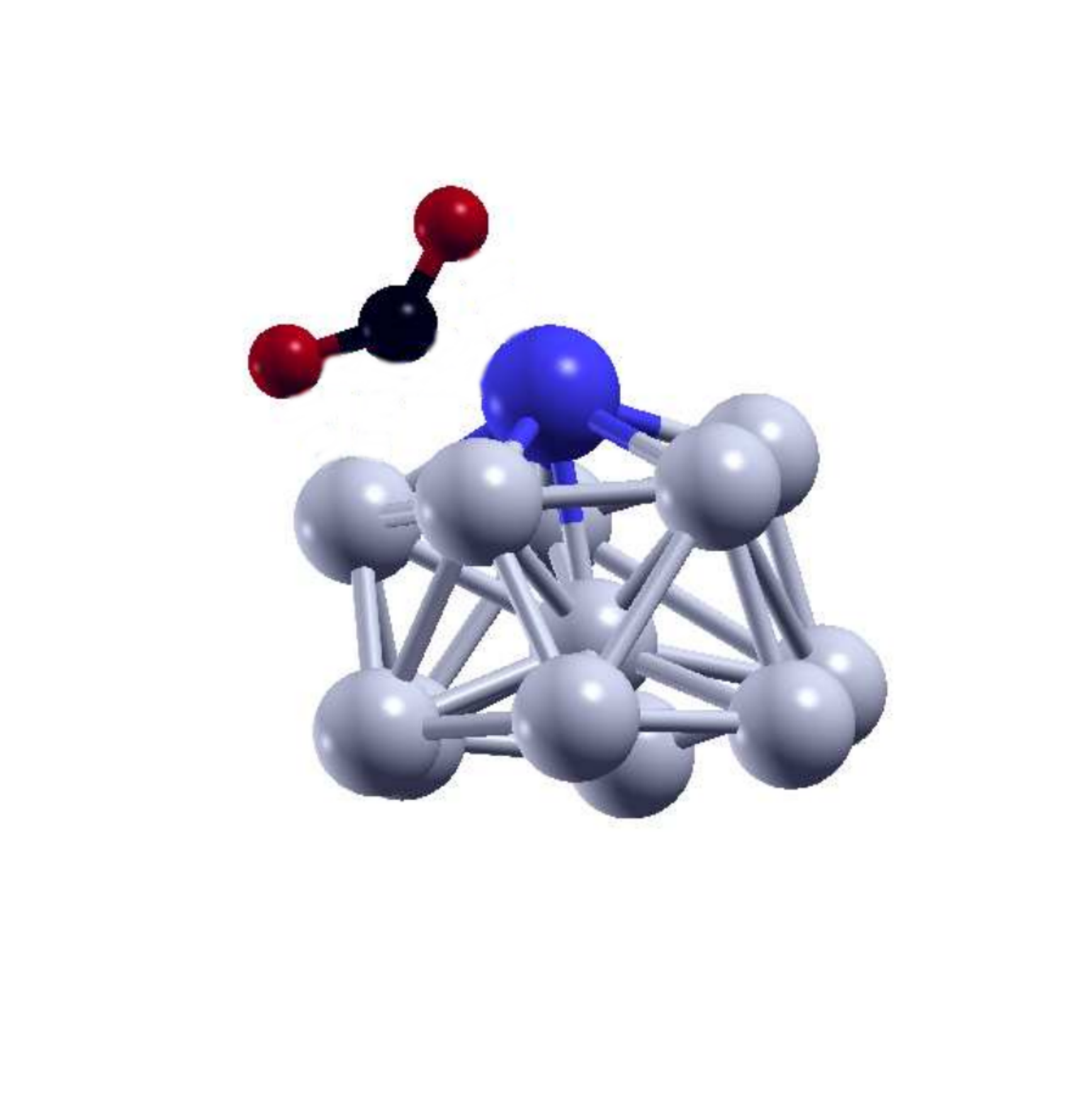} & \\\relax
 &&&&&\\
 \ce{[Al12Ru]-CO2} & 2.1 (C-Ru) & 142.9 & \ce{[Al12Co]-CO2} & 1.8 (C-Co) & 135.5 \\
 \hline
 &&&&&\\
  &\includegraphics[width=19mm,scale=0.08]{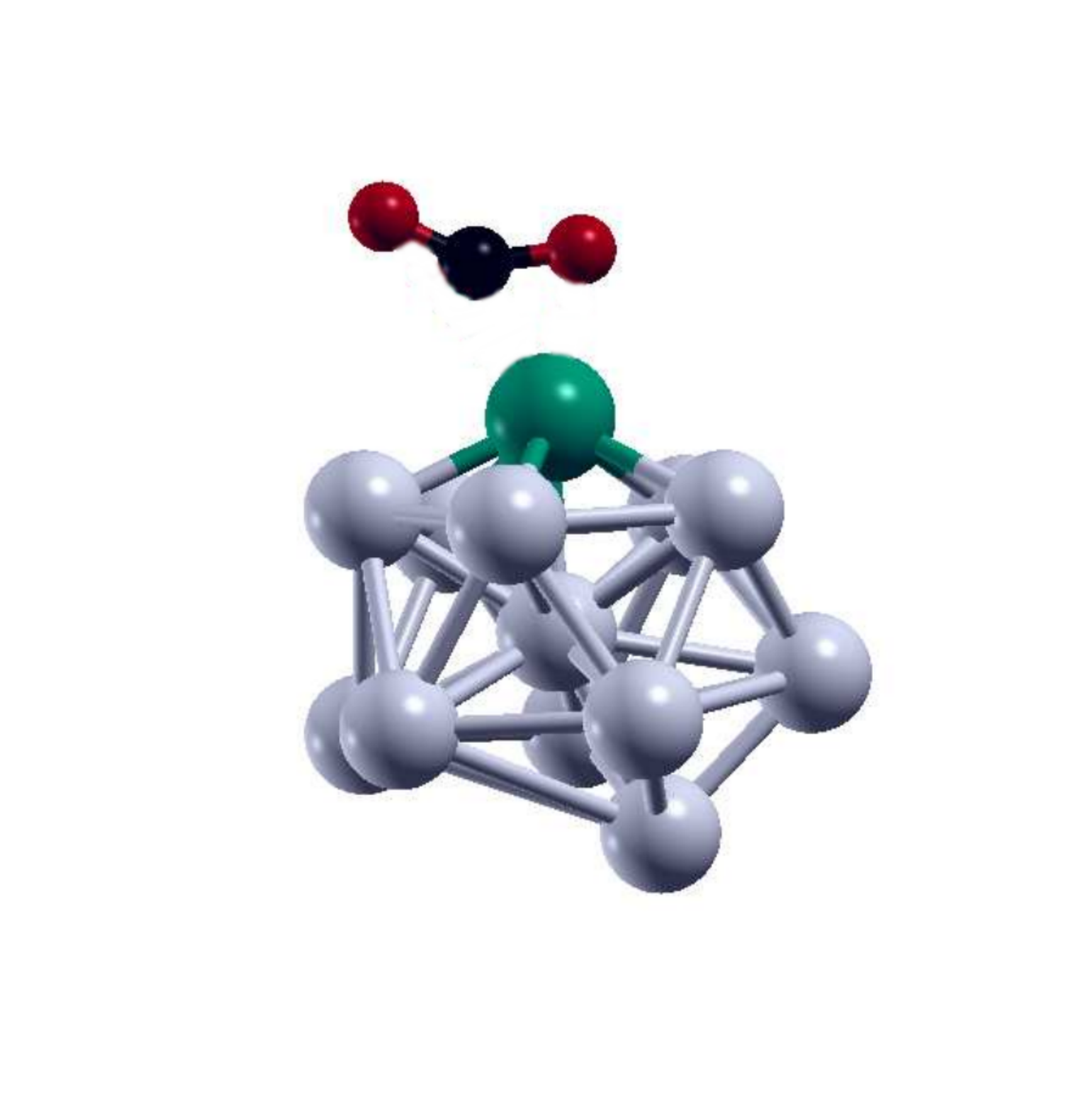} & && \includegraphics[width=19mm,scale=0.08]{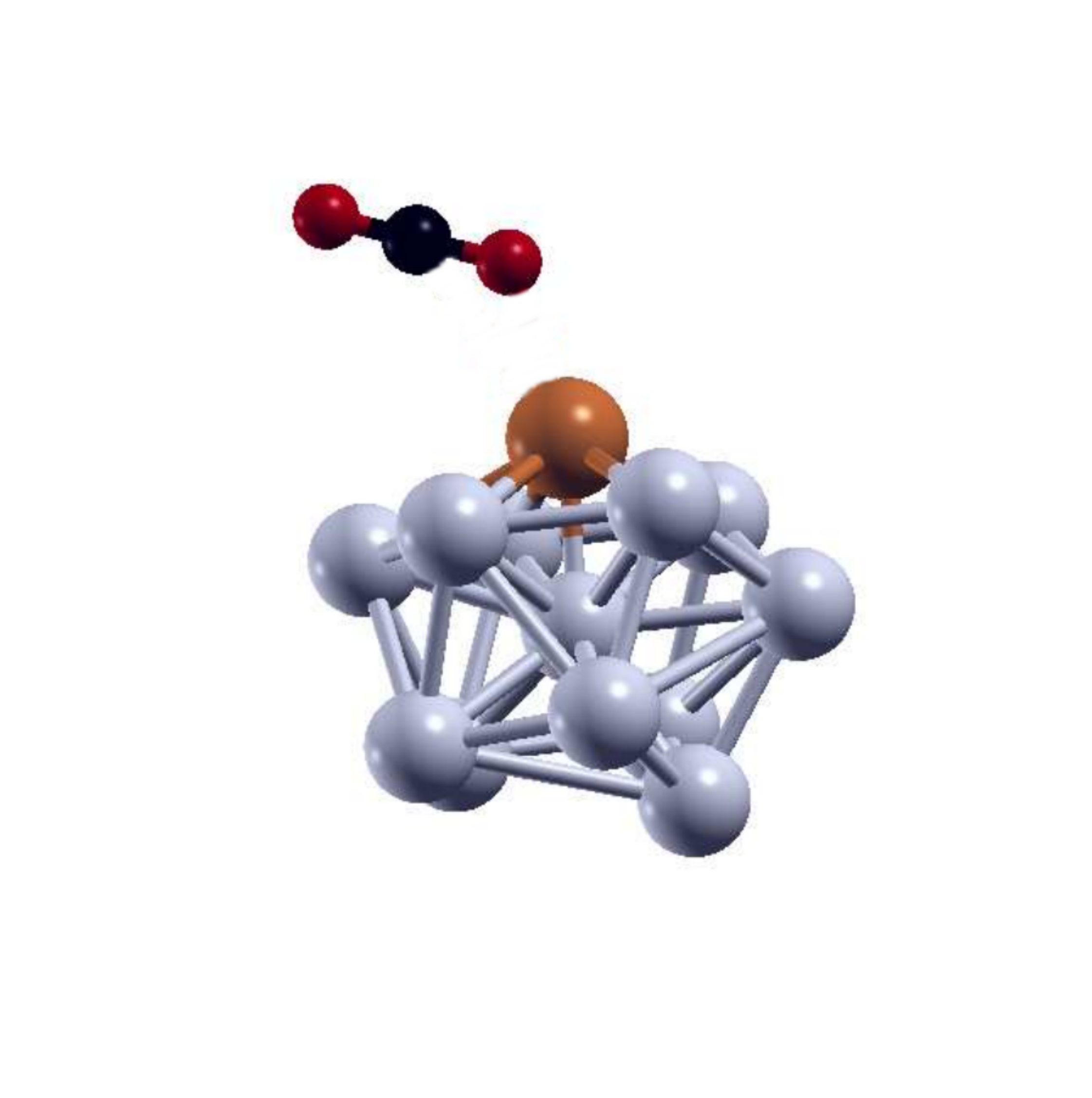} & \\\relax
  &&&&&\\
 \ce{[Al12Ni]-CO2} & 2.0 (O-Ni) & 146.2 & \ce{[Al12Cu]-CO2} & 2.3 (O-Cu) & 179.3 \\
\hline
\end{tabular}
\caption{Relaxed structures of \ce{[Al13]} and \ce{[Al12\textit{M}]} clusters with adsorbed \ce{CO2}. The adsorption distance is defined as the minimum distance between any molecular and any cluster atom (the relevant atoms are specified in brackets). The third column shows the O-C-O angle in degrees.}\label{gsdata geometry}
\end{table}

Table~\ref{electronicData} summarizes the electronic ground state properties of the various clusters with and without \ce{CO2}. As expected, systems with small adsorption distance have a large adsorption energy defined as the total energy difference of the relaxed system and its isolated ingredients (with each system being in its lowest-energy spin state). For example, \ce{[Al12Zr]-CO2} and \ce{[Al12Co]-CO2} (both with an adsorption distance of \SI{1.8}{\angstrom}) have an adsorption energy of -2.0~eV, while \ce{[Al12Fe]-CO2} and \ce{[Al12Cu]-CO2} both have adsorption energies of -0.3~eV. 

In all systems except \ce{[Al12Fe]-CO2} and \ce{[Al12Cu]-CO2}, adsorption is accompanied by a significant charge transfer between cluster and \ce{CO2}. Using a Hirshfeld analysis\cite{hirshfel1977}, we find that up to 0.36 electrons are transferred into the \ce{CO2}. The largest charge transfer is found in \ce{[Al12Zr-CO2]} which has an adsorption distance of \SI{1.9}{\angstrom} and one of the highest adsorption energies of -2.0~eV. Interestingly, systems with small adsorption energies (\ce{[Al12Fe]-CO2} and \ce{[Al12Cu]-CO2}) have a small charge transfer in the opposite directions, i.e. electrons are transferred from the \ce{CO2} onto the cluster.

\begin{figure}[t!]
\graphicspath{ {./ELFs/} }
  \includegraphics[width=210mm,scale=3]{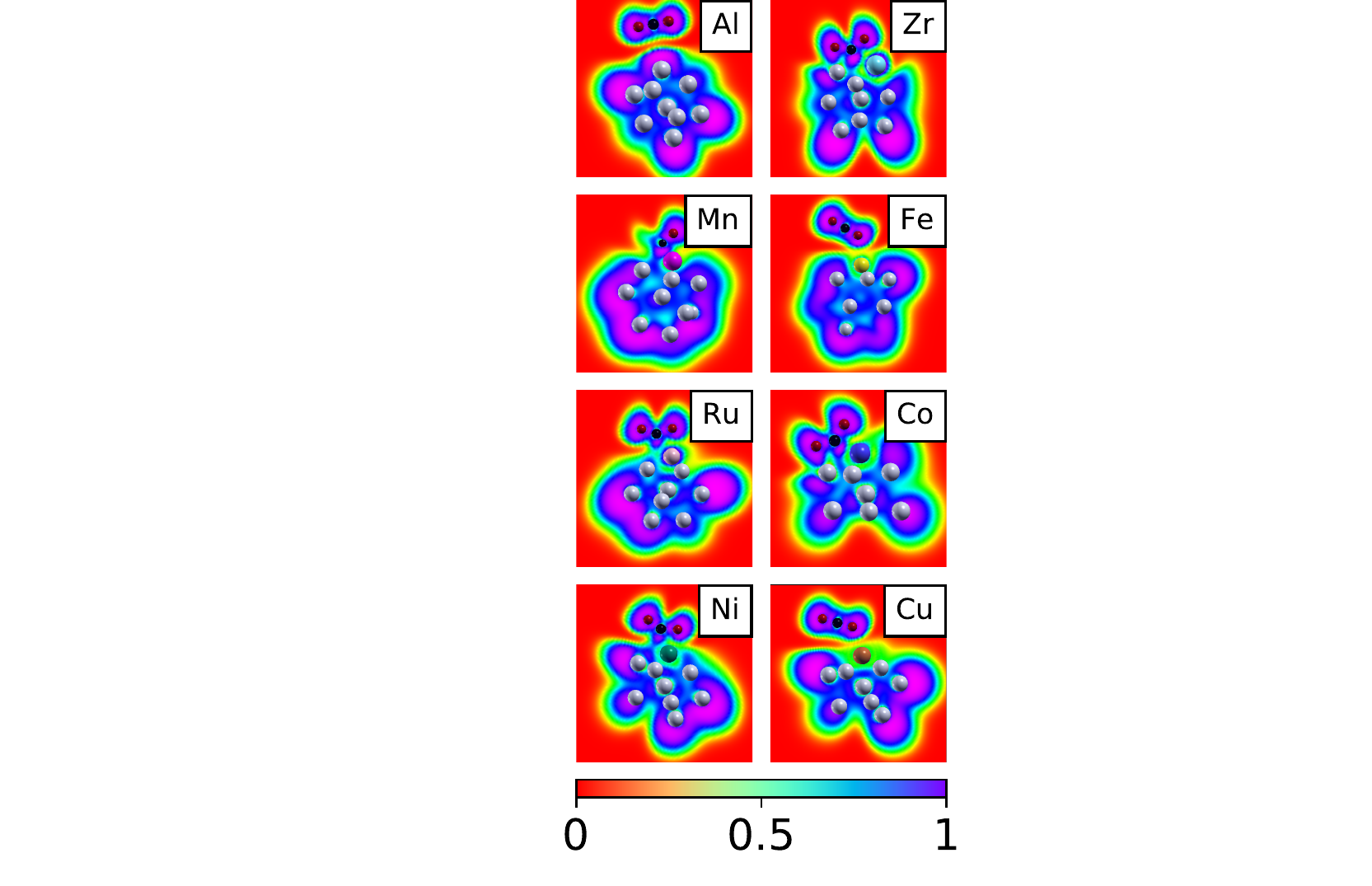}
\caption{Electron localization function (ELF) of the transition-metal doped Al clusters with \ce{CO2}. For comparison, the ELF of a neutral \ce{[Al13]} cluster with \ce{CO2} is also shown (upper left picture).}
\label{ELFs}
\end{figure}

Further insight into the ground state electronic structure can be obtained from the electron localization function (ELF), see Figure~\ref{ELFs}. The ELF is a qualitative measure for the localization character of electrons.\cite{becke1990simple} In all systems which exhibit electron transfer onto the \ce{CO2} molecule, an additional lobe in the ELF appears at the carbon atom with the lobe pointing towards the cluster. The value of the ELF in the lobe is close to unity indicating the presence of a highly localized electron. This is consistent with the expectation for a negatively charged \ce{CO2} molecule where the presence of an additional electron results in a so-called "Umpolung" of the carbon dioxide molecule, i.e. the partial charge of the carbon changes sign compared to the neutral molecule.\cite{alvarez2017co2} Comparing the ELFs of the transition metal doped clusters with adsorbed \ce{CO2} to the undoped \ce{[Al13]-CO2}, we find that the most important changes to the electronic structure are localized to the region near the dopant.

Table~\ref{electronicData} also shows the spin states and the vertical and adiabatic ionization potentials (vertical IPs were calculated using the Delta-SCF method, adiabatic ones were obtained from the difference in total energies of the relaxed ionized system and that of the relaxed neutral system) of the cluster with \ce{CO2} and compares them to the corresponding values of the isolated clusters without \ce{CO2}. For the isolated clusters, we find that \ce{[Al12Zr]}, \ce{[Al12Ru]} and \ce{[Al12Ni]} have a singlet ground state and that \ce{[Al12Cu]} has a doublet ground state. In contrast, high-spin ground states are obtained for \ce{[Al12Mn]} (sextet), \ce{[Al12Fe]} (quintet) and \ce{[Al12Co]} (quartet). Upon addition of the \ce{CO2}, only the spin state of the \ce{[Al12Co]} and the \ce{[Al12Mn]} system changes with \ce{[Al12Co-CO2]} and \ce{[Al12Mn-CO2]} both being a doublet. Both of these systems have a very small adsorption distance and exhibit a large electron transfer into the \ce{CO2}.\newline
The vertical ionization energies of the isolated clusters range from 6.2~eV (for \ce{[Al12Zr]}, \ce{[Al12Mn]}, \ce{[Al12Fe]}, \ce{[Al12Ru]}) to 6.5~eV in \ce{[Al12Cu]}. In all systems except for \ce{[Al12Fe]} and \ce{[Al12Cu]}, addition of the \ce{CO2} results in an increase of the ionization potential corresponding to an electronic stabilization of the system. This can be explained by the Nephelauxetic effect\cite{konig1971nephelauxetic}: For systems with ground-state charge-transfer, the electron is localized over hybridized orbitals formed by the cluster and the molecule which reduces electron-electron repulsion. This also explains why the Mn and the Co doped clusters change their spin-multiplicity into a low-spin state upon \ce{CO2} adsorption. The adiabatic ionization energies of the adsorbed systems are slightly smaller than their vertical counterparts for all systems except for \ce{[Al12Ru]-CO2} where it is the same. It is worth noting that relaxation after ionization leads to strong distortions in \ce{[Al12Fe]-Co2}, \ce{[Al12Mn]-Co2} and \ce{[Al12Ni]-Co2}, resulting in a more flattened geometry and in the case of the Ni doped cluster in an increase of adsorption distance and a change in \ce{CO2} bond angle which results in a linear molecule. The spin multiplicity of the ionized systems is equal to the multiplicity of the neutral system -1 in all cases except for \ce{([Al12Mn]-CO2)+}. Instead of being a singlet, the resulting spin state is a triplet.\newline
Comparison of the vertical with the adiabatic ionization potentials of the isolated clusters shows a decrease in ionization energy of 0.4~eV for the isolated Cu-doped cluster while all other clusters only exhibit minor changes. The main difference between the adiabatic ionization potentials and the vertical ones is the fact that from the adiabatic ionization potentials, the Cu-doped cluster unlike its Fe-doped counter-part does not get destabilized upon \ce{CO2} adsorption but stabilized like the systems that show \ce{CO2} activation. For the isolated clusters, the relaxed ionized clusters all have the spin multiplicity of the neutral isolated cluster -1 except for the Co doped cluster which results in a singlet spin state instead of the expected triplet.

\begin{table}
\begin{tabular}{ p{2.8cm} p{3.3cm} p{2.2cm} p{1.8cm} p{1.8cm} p{1.8cm} }
\hline
System & Spin multiplicity & Vertical ionization energy [eV] & Adiabatic ionization energy [eV] & Adsorption Energy [eV] & GS CT into \ce{CO2} [e]\\
\hline
\ce{[Al12Zr]-CO2} & singlet (singlet)  & 6.3 (6.2)& 6.1 (6.0)& -2.0 & +0.36\\\relax
\ce{[Al12Mn]-CO2} & doublet (sextet) & 6.5 (6.2)&6.3 (6.0) &-1.5 & +0.28\\\relax
\ce{[Al12Fe]-CO2} & quintet (quintet) & 6.0 (6.2)& 5.8 (6.1)&-0.3 & -0.08\\\relax
\ce{[Al12Ru]-CO2} & singlet (singlet) & 6.4 (6.2)& 6.4 (6.2)&-1.2 & +0.31\\\relax
\ce{[Al12Co]-CO2} & doublet (quartet) & 6.4 (6.3)& 6.3 (6.2)&-2.0 & +0.29\\\relax
\ce{[Al12Ni]-CO2} & singlet (singlet) & 6.7 (6.4)& 6.5 (6.3)&-0.7 & +0.20\\\relax
\ce{[Al12Cu]-CO2} & doublet (doublet) & 6.4 (6.5) & 6.3 (6.1)&-0.3 & -0.07\\
\hline
\end{tabular}
\caption{Electronic ground-state properties of \ce{[Al12\textit{M}]} clusters with adsorbed \ce{CO2}, including the spin multiplicity, ionization energies, adsorption energy and the ground-state charge transfer (GS CT) into \ce{CO2} (evaluated using Hirshfeld's scheme\cite{hirshfel1977}). Values for the isolated clusters are given in brackets.}\label{electronicData}
\end{table}

\begin{figure}[t!]
\graphicspath{ {./pDOS/} }
  \includegraphics[width=70mm,scale=0.5]{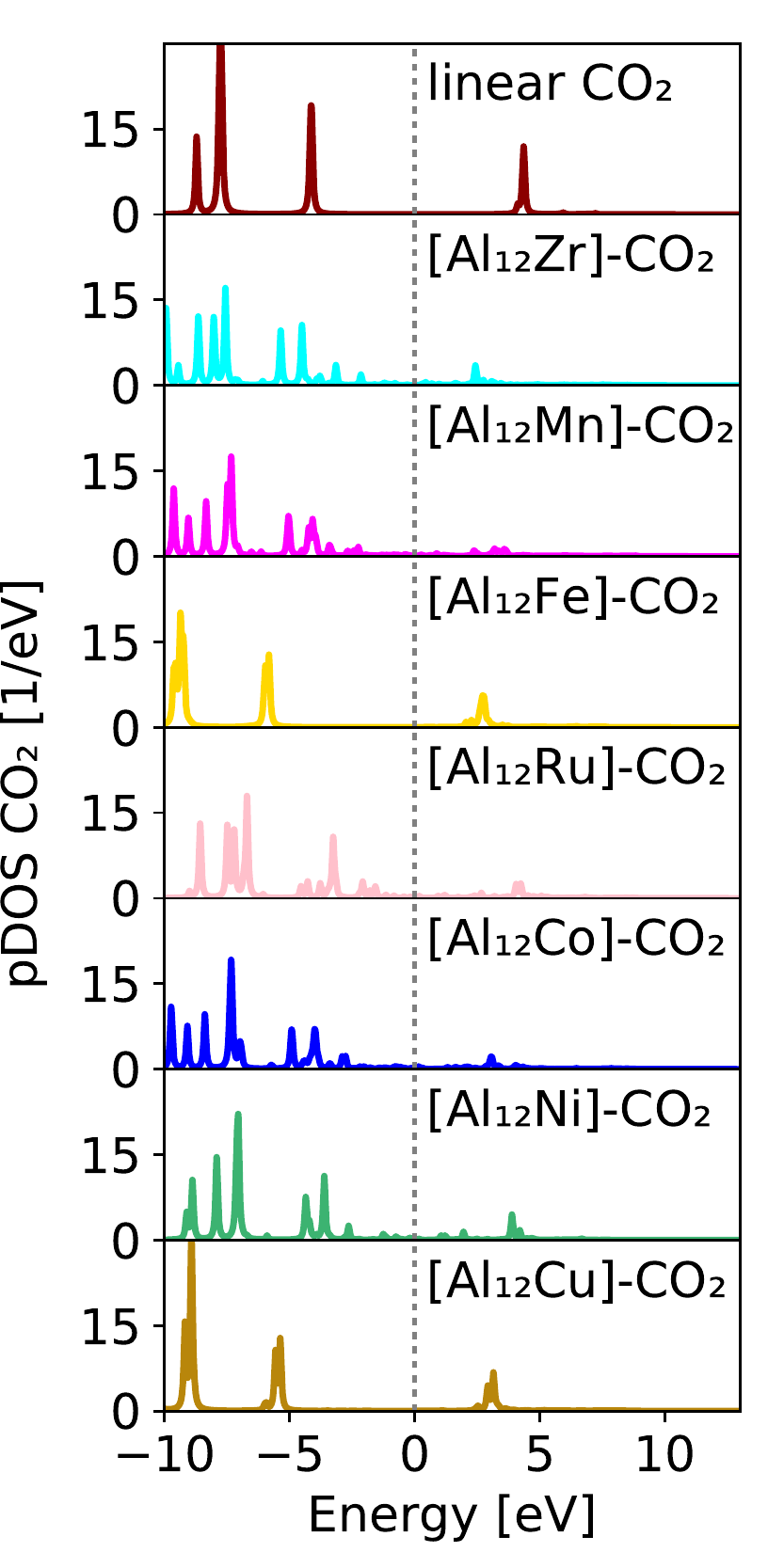}
\caption{DOS projected onto \ce{CO2} atomic orbitals for transition metal doped aluminium cluster with adsorbed \ce{CO2}. The zero of energy is set to the Fermi level and denoted by a grey vertical line. A broadening of 0.05~eV has been used. For comparison, we also show the DOS of an isolated \ce{CO2} molecule.}
\label{pDOS}
\end{figure}

Figure~\ref{pDOS} shows the density of states (DOS) of all systems projected onto the \ce{CO2} molecule. Two different scenarios are found: if the \ce{CO2} molecule is weakly bound to the cluster (as in the case of \ce{[Al12Cu]-CO2} and \ce{[Al12Fe]-CO2}), the projected DOS (pDOS) resembles that of the isolated molecule. In particular, the peak above the Fermi energy results from the lowest unoccupied molecular orbital (LUMO) of the \ce{CO2} while the peak at $\sim - 6$~eV corresponds to the \ce{CO2} HOMO. There are no peaks in the vicinity of the Fermi level. In contrast, the pDOS of systems with strong \ce{CO2} adsorption (all systems except \ce{[Al12Cu]-CO2} and \ce{[Al12Fe]-CO2}) exhibits a large number of peaks as a consequence of hybridization between molecular states and cluster states. Compared to the systems with weak adsorption, we find occupied states with partial \ce{CO2} character relatively close to the Fermi level. These states contain the extra electrons that appear on \ce{CO2} in the Hirshfeld analysis as a result of hybridisation between occupied cluster and unoccupied molecular orbitals, see Table~\ref{electronicData}.

\begin{figure}[t!]
\graphicspath{ {./spectra/} }
  \includegraphics[width=70mm,scale=0.5]{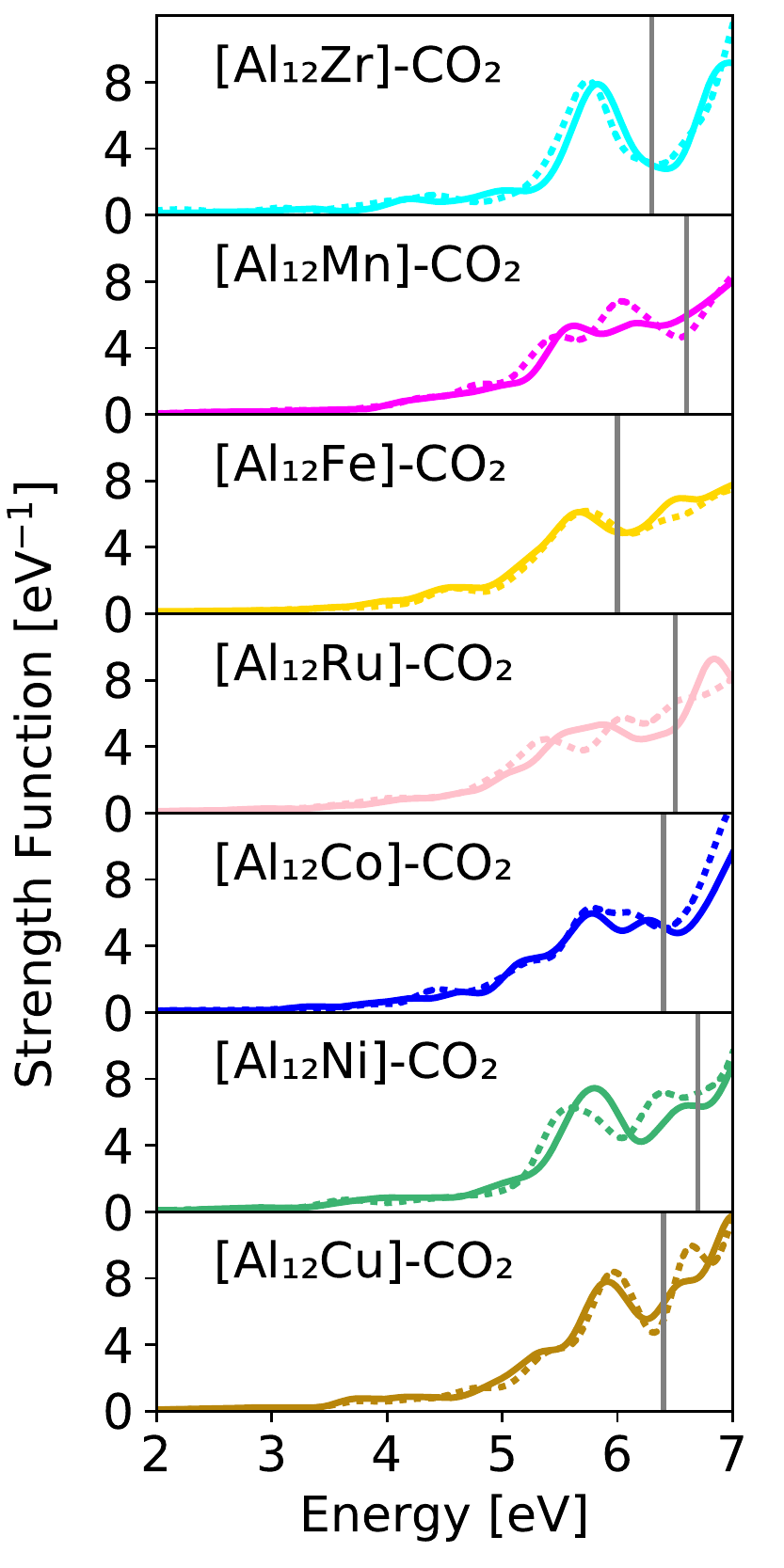}
\caption{Absorption spectra of the transition metal doped Al clusters with (solid lines) and without (dashed lines) adsorbed \ce{CO2}. The strength function is proportional to the absorption cross section.\cite{marques2003octopus} The vertical grey line denotes the first vertical ionization energy of each system.}
\label{spectra-tp}
\end{figure}

Next, we consider excited states of the transition metal doped Al clusters with \ce{CO2} molecules. Fig.~\ref{spectra-tp} shows the absorption spectra of all systems with and without the \ce{CO2}. All spectra have an onset between 3 and 4~eV and exhibit a peak between 5 and 6~eV with the detailed shape of the spectrum depending on the transition metal dopant and the cluster geometry. In \ce{[Al12Cu]-CO2} and \ce{[Al12Fe]-CO2}, the presence of the \ce{CO2} molecule does not modify the absorption spectrum significantly. This is not surprising as the \ce{CO2} molecule has a large HOMO-LUMO gap (eigenenergy(\ce{CO2}LUMO) - eigenenergy(\ce{CO2}HOMO) = 8.5~eV according to our calculations and the findings of Kim et al.\cite{kim2018energy}) and the interaction between cluster and molecule is weak. In the other systems, small differences can be observed, however. For example, the presence of the \ce{CO2} results in a noticeable blue-shift of the main absorption peak near 5.5~eV in \ce{[Al12Ni]-CO2}.

\begin{figure}[t!]
\graphicspath{ {./LICT/} }
  \includegraphics[width=70mm,scale=0.5]{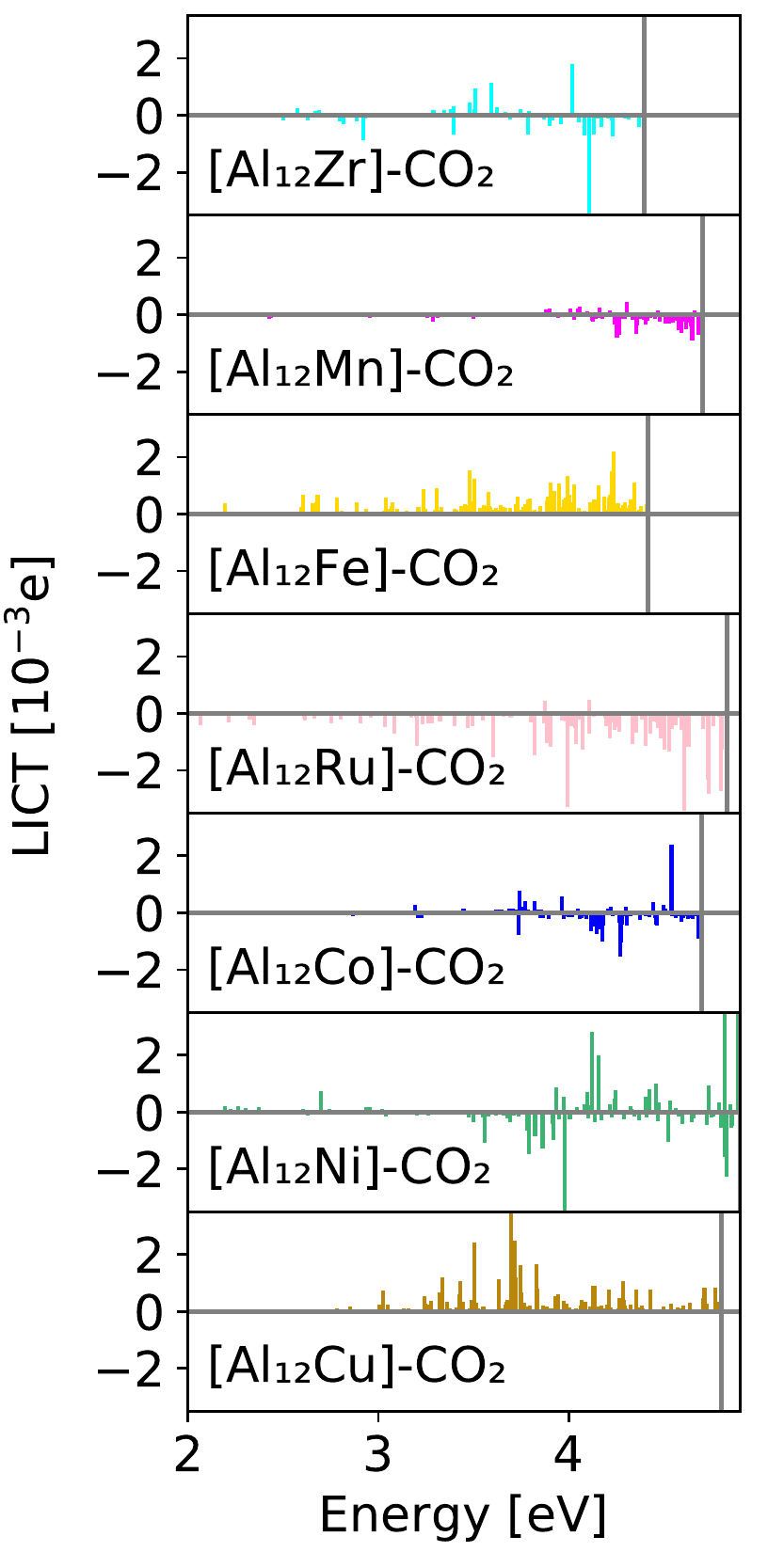}
\caption{Light-induced charge transfer (LICT) in units of elementary charge e as function of excitation energy between transition metal doped Al clusters and \ce{CO2}. Grey vertical lines indicate the threshold energy above which the excited states contain a significant fraction (more than five percent) of Kohn-Sham orbitals above the vacuum level.}
\label{LICT}
\end{figure}

Finally, we investigate the light-induced charge transfer between the clusters and the \ce{CO2}. Fig.~\ref{LICT} shows the amount of transferred charge as function of the excitation energy. First, we observe that \ce{[Al12Cu]} and \ce{[Al12Fe]} only transfer electrons into the \ce{CO2}. This is expected as there is no charge transfer in the ground state for these systems and the \ce{CO2} HOMO lies far below the Fermi level, see Fig.~\ref{pDOS}. As the \ce{CO2} LUMO is closer to the Fermi level in the \ce{[Al12Fe]-CO2} system compared to the \ce{[Al12Cu]-CO2} system, the onset of LICT occurs at somewhat lower excitation energies. In contrast, \ce{[Al12Ru]-CO2} and \ce{[Al12Mn]-CO2} exhibit mostly LICT in the opposite direction, i.e. electrons are transferred from the \ce{CO2} to the cluster. This can be interpreted as a back-transfer of those electrons that move onto the cluster in the ground state, see Table~\ref{electronicData}. Among these systems, the back-transfer is most pronounced in the Ru-doped cluster and already starts at low excitation energies of $\sim 2$~eV. Inspecting the pDOS of the Ru-cluster, Fig.~\ref{pDOS}, we see that this system exhibits hybridized occupied states close to the Fermi level. In the Mn-doped system, however, LICT is relatively small and only occurs at high excitation energies. This is caused by the fact that the Mn-doped cluster undergoes a transition from high-spin to a low-spin ground state when in contact with the \ce{CO2} (see Table~\ref{electronicData}), making a back-transfer less favorable than in the Ru-doped system where the spin state is not affected by the presence of the \ce{CO2}. A similar situation occurs for the Co-doped system. Finally, for \ce{[Al12Zr]-CO2}, \ce{[Al12Co]-CO2} and \ce{[Al12Ni]-CO2}, LICT can occur in both directions. Interestingly, \ce{[Al12Zr]-CO2} and \ce{[Al12Co]-CO2} exhibit only a small reduction of the ionization energy upon \ce{CO2} adsorption. This suggests that the transferred electrons in these systems originate from delocalized cluster orbitals such that additional electron transfer in the excited state does not result in a large energy cost due to repulsive electron-electron interactions.

\section{Discussion and conclusions}

We have studied the atomic and electronic properties of transition metal doped aluminium clusters with adsorbed \ce{CO2} using first-principles density-functional theory and also time-dependent density-functional theory. Our results suggest that the set of transition-metal doped aluminium clusters can be divided into two categories: (i) clusters which strongly adsorb \ce{CO2} with significant ground state charge transfer and (ii) clusters with weak \ce{CO2} adsorption and no significant ground state charge transfer. The clusters in class (ii) are \ce{[Al12Cu]} and \ce{[Al12Fe]}, while all other systems (\ce{[Al12Zr]}, \ce{[Al12Mn]}, \ce{[Al12Ru]}, \ce{[Al12Co]}, \ce{[Al12Ni]}) belong to class (i). Systems in class (ii) exhibit light-induced electron transfer into the \ce{CO2}. In contrast, light-induced electron back-transfer from the \ce{CO2} to the cluster dominates in class (i) systems, but light-induced electron transfer is also possible when the \ce{CO2} adsorption does not result in significant changes to the ionization energy (as in the cases of \ce{[Al12Zr]} and \ce{[Al12Co]}). In conclusion, our work contributes to a detailed understanding of the effect of transition metal doping on ground- and excited-state charge transfer and paves the way towards a rational design of atomically defined photo-catalysts for \ce{CO2} conversion.

\begin{acknowledgement}
This work was supported by the European Research Council (ERC-2015-AdG694097), the Cluster of Excellence ‘Advanced Imaging of Matter' (AIM), Grupos Consolidados (IT1249-19), SFB925 "Light induced dynamics and control of correlated quantum systems" and by the Deutsche Forschungsgemeinschaft (DFG) through the Research Training Group -Quantum Mechanical Materials Modelling -  GRK 2247. The Flatiron Institute is a division of the Simons Foundation.
\end{acknowledgement}

\bibliography{acs-achemso}

\title{Supporting Information for "Light Induced Charge Transfer from Transition-metal Doped Aluminium Clusters to Carbon Dioxide"}

\begin{suppinfo}
	
	Table~\ref{convergence_parameters_gs} and \ref{convergence_parameters_time-prop} summarize the convergence parameters that were used in this work.

	\begin{table}
		\begin{tabular} {p{2.8cm} p{1.8cm} p{1.8cm} p{1.8cm} p{1.8cm} p{1.8cm} p{1.8cm}}
			\hline
			System & Grid (IP) [\si{\angstrom}] & Box (IP)[\si{\angstrom}] & Grid (AE) [\si{\angstrom}] & Box (AE) [\si{\angstrom}] & Grid (DOS) [\si{\angstrom}] & Box (DOS) [\si{\angstrom}] \\
			\hline
			\ce{[Al12Zr]-CO2} & 0.18 & 5 & 0.06 & 5 & 0.16 & 9\\\relax
			\ce{[Al12Mn]-CO2} & 0.14 & 6  & 0.06 & 5 & 0.14 & 9\\\relax
			\ce{[Al12Fe]-CO2} & 0.18 & 5  & 0.06 & 5 & 0.16 & 9\\\relax
			\ce{[Al12Ru]-CO2} & 0.20 & 6  & 0.06 & 5 & 0.18 & 9\\\relax
			\ce{[Al12Co]-CO2} & 0.16 & 5  & 0.06 & 5 & 0.14 & 9 \\\relax
			\ce{[Al12Ni]-CO2} & 0.12 & 5  & 0.06 & 5 & 0.14 & 9\\\relax
			\ce{[Al12Cu]-CO2} & 0.16 & 5  & 0.06 & 5 & 0.14 & 9\\
			\hline
		\end{tabular}
		\caption{Calculation parameters used for determining the vertical ionization energy ("IP"), adsorption energy, adiabatic ionization energy and most favorable spin-state ("AE") and computation of DOS, pDOS, ELFs, ground-state electron transfer and Casida calculations ("DOS"). "Grid" denotes the spacing of points on the real-space grid while "Box" denotes the radius of the minimum box. In the calculation of the ELF of \ce{[Al13]-CO2}, a smearing of 0.1~eV was used to achieve convergence.}\label{convergence_parameters_gs}
	\end{table}

	\begin{table}
		\begin{tabular} {p{2.8cm} p{2.3cm} p{2.3cm} p{2.3cm} p{2.3cm}}
			\hline
			System   &  Time Step [1/eV] & Grid [\si{\angstrom}] & Box Radius [\si{\angstrom}] & kick strength [1/\si{\angstrom}]\\
			\hline
			\ce{[Al12Zr]-CO2} & 0.001 & 0.26 & 8.0 &  0.01\\\relax
			\ce{[Al12Mn]-CO2} & 0.001 & 0.18 & 8.0 & 0.01\\\relax
			\ce{[Al12Fe]-CO2} & 0.001 & 0.22 & 8.0 & 0.01 \\\relax
			\ce{[Al12Ru]-CO2} & 0.001 & 0.32 & 8.0 & 0.01\\\relax
			\ce{[Al12Co]-CO2} & 0.001 & 0.20 & 9.0 & 0.01\\\relax
			\ce{[Al12Ni]-CO2} & 0.001 & 0.18 & 8.0 & 0.01\\\relax
			\ce{[Al12Cu]-CO2} & 0.001 & 0.18 & 13.0 & 0.01\\
			\hline
		\end{tabular}
		\caption{Minimal convergence Parameters found for computation of the absorption spectra from time-propagations.}\label{convergence_parameters_time-prop}
	\end{table}

\end{suppinfo}

\begin{figure}[t!]
	\graphicspath{ {./TOC/} }
	\includegraphics[scale=1.0]{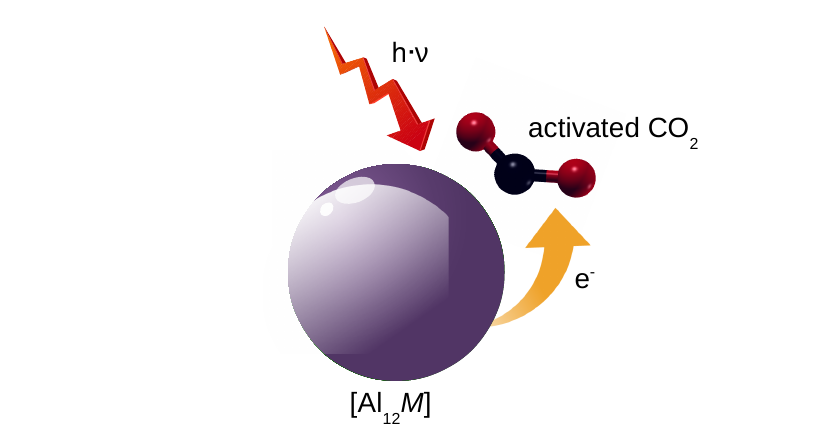}
	\caption{For Table of Contents Only.}
	\label{TOC}
\end{figure}

\end{document}


Table~\ref{convergence_parameters_gs} and \ref{convergence_parameters_time-prop} summarize the convergence parameters that were used in this work.

\begin{table}
 \begin{tabular} {p{2.8cm} p{1.8cm} p{1.8cm} p{1.8cm} p{1.8cm} p{1.8cm} p{1.8cm}}
 \hline
 System & Grid (IP) [Å] & Box (IP)[Å] & Grid (AE) [Å] & Box (AE) [Å] & Grid (DOS) [Å] & Box (DOS) [Å] \\
 \hline
 \ce{[Al12Zr]-CO2} & 0.18 & 5 & 0.06 & 5 & 0.16 & 9\\\relax
 \ce{[Al12Mn]-CO2} & 0.14 & 6  & 0.06 & 5 & 0.14 & 9\\\relax
 \ce{[Al12Fe]-CO2} & 0.18 & 5  & 0.06 & 5 & 0.16 & 9\\\relax
 \ce{[Al12Ru]-CO2} & 0.20 & 6  & 0.06 & 5 & 0.18 & 9\\\relax
 \ce{[Al12Co]-CO2} & 0.16 & 5  & 0.06 & 5 & 0.14 & 9 \\\relax
 \ce{[Al12Ni]-CO2} & 0.12 & 5  & 0.06 & 5 & 0.14 & 9\\\relax
 \ce{[Al12Cu]-CO2} & 0.16 & 5  & 0.06 & 5 & 0.14 & 9\\
\hline
\end{tabular}
\caption{Calculation parameters used for determining the vertical ionization energy ("IP"), adsorption energy, adiabatic ionization energy and most favorable spin-state ("AE") and computation of DOS, pDOS, ELFs, ground-state electron transfer and Casida calculations ("DOS"). "Grid" denotes the spacing of points on the real-space grid while "Box" denotes the radius of the minimum box. In the calculation of the ELF of \ce{[Al13]-CO2}, a smearing of 0.1~eV was used to achieve convergence.}\label{convergence_parameters_gs}
\end{table}

\begin{table}
\begin{tabular} {p{2.8cm} p{2.3cm} p{2.3cm} p{2.3cm} p{2.3cm}}
 \hline
 System   &  Time Step [1/eV] & Grid [Å] & Box Radius [Å] & kick strength [1/Å]\\
 \hline
 \ce{[Al12Zr]-CO2} & 0.001 & 0.26 & 8.0 &  0.01\\\relax
 \ce{[Al12Mn]-CO2} & 0.001 & 0.18 & 8.0 & 0.01\\\relax
 \ce{[Al12Fe]-CO2} & 0.001 & 0.22 & 8.0 & 0.01 \\\relax
 \ce{[Al12Ru]-CO2} & 0.001 & 0.32 & 8.0 & 0.01\\\relax
 \ce{[Al12Co]-CO2} & 0.001 & 0.20 & 9.0 & 0.01\\\relax
 \ce{[Al12Ni]-CO2} & 0.001 & 0.18 & 8.0 & 0.01\\\relax
 \ce{[Al12Cu]-CO2} & 0.001 & 0.18 & 13.0 & 0.01\\
\hline
\end{tabular}
\caption{Minimal convergence Parameters found for computation of the absorption spectra from time-propagations.}\label{convergence_parameters_time-prop}
\end{table}